\documentclass[conference, onecolumn]{IEEEtran}
\IEEEoverridecommandlockouts
\usepackage{cite}
\usepackage{bm}
\usepackage{amsmath,amssymb,amsfonts}
\usepackage{algorithm}
\usepackage{algorithmic}
\usepackage{ifpdf}
\ifCLASSINFOpdf
\usepackage[pdftex]{graphicx}
\DeclareGraphicsExtensions{.pdf,.jpeg,.png}
\else
\usepackage[dvips]{graphicx}
\DeclareGraphicsExtensions{.eps}
\fi
\graphicspath{{./Figures/}}
\usepackage{graphicx}
\ifCLASSOPTIONcompsoc
\usepackage[caption=false, font=scriptsize, labelfont=sf, textfont=sf]{subfig}
\else
\usepackage[caption=false, font=footnotesize]{subfig}
\fi
\usepackage{epstopdf}
\usepackage[letterpaper, left=0.625in, right=0.625in, bottom=1in, top=0.70in]{geometry}

\usepackage{textcomp}
\usepackage{xcolor}

\newcommand{\be}{\begin{equation}}
\newcommand{\ee}{\end{equation}}

\def\BibTeX{{\rm B\kern-.05em{\sc i\kern-.025em b}\kern-.08em
    T\kern-.1667em\lower.7ex\hbox{E}\kern-.125emX}}

\begin{document}

\title{CSMAAFL: Client Scheduling and Model Aggregation in Asynchronous Federated Learning
}
\author{\IEEEauthorblockN{Xiang Ma\IEEEauthorrefmark{1}, Qun Wang\IEEEauthorrefmark{2}, Haijian Sun\IEEEauthorrefmark{3}, Rose Qingyang Hu\IEEEauthorrefmark{1}, and Yi Qian\IEEEauthorrefmark{4} }
\IEEEauthorblockA{
\IEEEauthorrefmark{1}Department of Electrical and Computer Engineering, Utah State University, Logan, UT \\
\IEEEauthorrefmark{2}Department of Computer Science, San Francisco State University, San Francisco, CA \\
\IEEEauthorrefmark{3}School of Electrical and Computer Engineering, University of Georgia, Athens, GA \\
\IEEEauthorrefmark{4}Department of Electrical and Computer Engineering, University of Nebraska–Lincoln, Omaha, NE
}
}
\maketitle

\begin{abstract}

Asynchronous federated learning aims to solve the straggler problem in heterogeneous environments, i.e., clients have small computational capacities that could cause aggregation delay. The principle of asynchronous federated learning is to allow the server to aggregate the model once it receives an update from any client rather than waiting for updates from multiple clients or waiting a specified amount of time in the synchronous mode. Due to the asynchronous setting, the stale model problem could occur, where the slow clients could utilize an outdated local model for their local data training. Consequently, when these locally trained models are uploaded to the server, they may impede the convergence of the global training.
Therefore, effective model aggregation strategies play a significant role in updating the global model. Besides, client scheduling is also critical when heterogeneous clients with diversified computing capacities are participating in the federated learning process. This work first investigates the impact of the convergence of asynchronous federated learning mode when adopting the aggregation coefficient in synchronous mode. The effective aggregation solutions that can achieve the same convergence result as in the synchronous mode are then proposed, followed by an improved aggregation method with client scheduling. The simulation results in various scenarios demonstrate that the proposed algorithm converges with a similar level of accuracy as the classical synchronous federated learning algorithm but effectively accelerates the learning process, especially in its early stage.

\end{abstract}


\section{Introduction}

The distributed learning nature of Federated Learning (FL) can effectively address the privacy concerns associated with machine learning by sharing only the refined model instead of exposing raw data to other devices. Under the coordination of the central server, clients collaboratively train a global model through an iterative process. 
The training process is organized into discrete rounds, with each round building upon the training results obtained in the previous round. This classical synchronous federated learning (SFL) could suffer straggler issues caused by slow clients \cite{straggler}.


 With the rapid advancement of pervasive intelligence, a wide range of heterogeneous devices can now serve as clients for FL. These devices include desktop computers, laptops, smartphones, Raspberry Pis, and more. By collecting data from nearby sensors, these devices can conduct local training and learning. However, computing resource-constraint devices, such as Raspberry Pis, may experience delays and become stragglers when processing substantial volumes of data. To address these straggler issues, various strategies have been proposed to be integrated within SFL frameworks.

In \cite{fed_avg}, a method was proposed where a small subset of clients is sampled in each round, eliminating the need to wait for updates from all clients. However, it does not fully address the straggler issue, as slower clients can still be selected within the subset.
A predefined synchronous window was suggested to aggregate as many client updates as possible in each round in \cite{fed_window}. However, slow clients may not get the opportunity to upload their updates during the entire learning process. Furthermore, it does not ensure the convergence of the model.
An adaptive local computation scheme in resource-constrained edge computing systems was proposed in \cite{fed_edge}, where fast clients can execute more local iterations than slow clients. Nevertheless, the global model still needs to wait for all clients to complete their local computations before getting the updates. 

Another strategy to tackle the straggler problem is called asynchronous federated learning (AFL) \cite{asyn_geo}, in which the process of local model uploading at clients and global aggregation at the server are separated. Thus aggregation of the global model is executed without the need to wait for receiving all the client models.
\cite{asyn_online} presented an online AFL algorithm with data being non-Independent and Identically Distributed (non-IID). 
The server initiates the aggregation process upon receiving the update from a single client. The newly aggregated model is then distributed to the clients deemed `ready'. However, there is a lack of definition of `ready' clients, nor does it outline a methodology on how to select.
Similarly, \cite{asyn_opt} proposed to decouple scheduling and aggregation but did not define the criteria for client selections in the scheduling. A Euclidean distance-based adaptive federated aggregation algorithm was introduced in \cite{asyn_ed} to solve the stale model problem in AFL. 
The staleness is measured using the distance between the current global model and the stale global model. This evaluation process requires the server to store all the global models, starting from the initial training phase up to the current iteration, leading to significant consumption of storage resources on the server.
\cite{asyn_wireless} considered AFL over wireless networks where a global model is broadcast to all the clients in each global iteration. Despite the various works mentioned above, few of them have offered a comprehensive architectural outline of AFL. Moreover, no comparative analysis between SFL and AFL has been provided.

In this work, a new AFL framework with client scheduling and model aggregation is developed. The proposed scheduling mechanism considers both computational capabilities and fairness. The model aggregation component is designed to address the staleness problem inherent in AFL. Besides, a detailed AFL architecture and a comparative analysis between AFL and SFL are given.
The major contributions of this paper are summarized as follows:
\begin{itemize}
    \item A new AFL architecture is introduced. In each global iteration, one client is selected to upload its local model. Subsequently, the newly aggregated global model is sent back only to the client that has just completed its local model update. 
    
    \item A comparative analysis of the completion time between AFL and SFL is presented. While the total learning completion time is not necessarily shorter in AFL, its distinct advantage lies in the timely updating of the global model within a significantly reduced timeframe. 
    
    \item A general baseline AFL framework is first introduced, which can achieve the same accuracy performance as SFL. Then, the solution of the aggregation coefficients is developed based on this framework setting. 
    
    \item A new AFL framework that incorporates both client scheduling and model aggregation is then proposed, which takes into account client computational capabilities, fairness, and model staleness as import metrics. As a result, the proposed AFL framework provides a solution to the inherent stale model issue. 
\end{itemize}

The rest of the paper is organized as follows. Section II introduces the system model of SFL and AFL, where the comparison of the two models is given. In Section III, the baseline AFL framework that can achieve the same learning performance as SFL is first presented. The advanced client scheduling and model aggregation framework is then developed. Simulation results are given in Section IV. The paper is concluded in Section V.

\section{System Model}
Classical federated averaging (FedAvg) algorithm is a synchronous communication model where the server performs aggregation after receiving models from a predetermined number of clients or after a set amount of time has elapsed. While in an asynchronous setting, the server commences model aggregation immediately upon receipt of an update. This ensures that the server is always updated with the most recent model.

In our work, the considered FL system is comprised of a central server and $M$ clients. Each client $m$ has $D_m$ amount of dataset. 
\subsection{Synchronous Federated Learning}

In the SFL framework, the learning process unfolds iteratively between the server and clients. Each iteration consists of four fundamental steps. Firstly (S1), the server disseminates the current global model to all clients. Secondly (S2), clients utilize the global model as the initial point and employ an optimization method such as stochastic gradient descent (SGD) to derive a new local model. Following this, in step 3 (S3), the updated local model is uploaded to the server. Finally (S4), the server awaits either a fixed amount of time or model updates from a predetermined number of clients before performing aggregation. This process can be observed in Figure 1 (left). Notably, in SFL, a ``wait'' stage is present to allow all clients to upload their respective local models, thereby preventing the server from aggregating prematurely.

\begin{figure}[!th]
	\centering
	\includegraphics[width=0.7\linewidth]{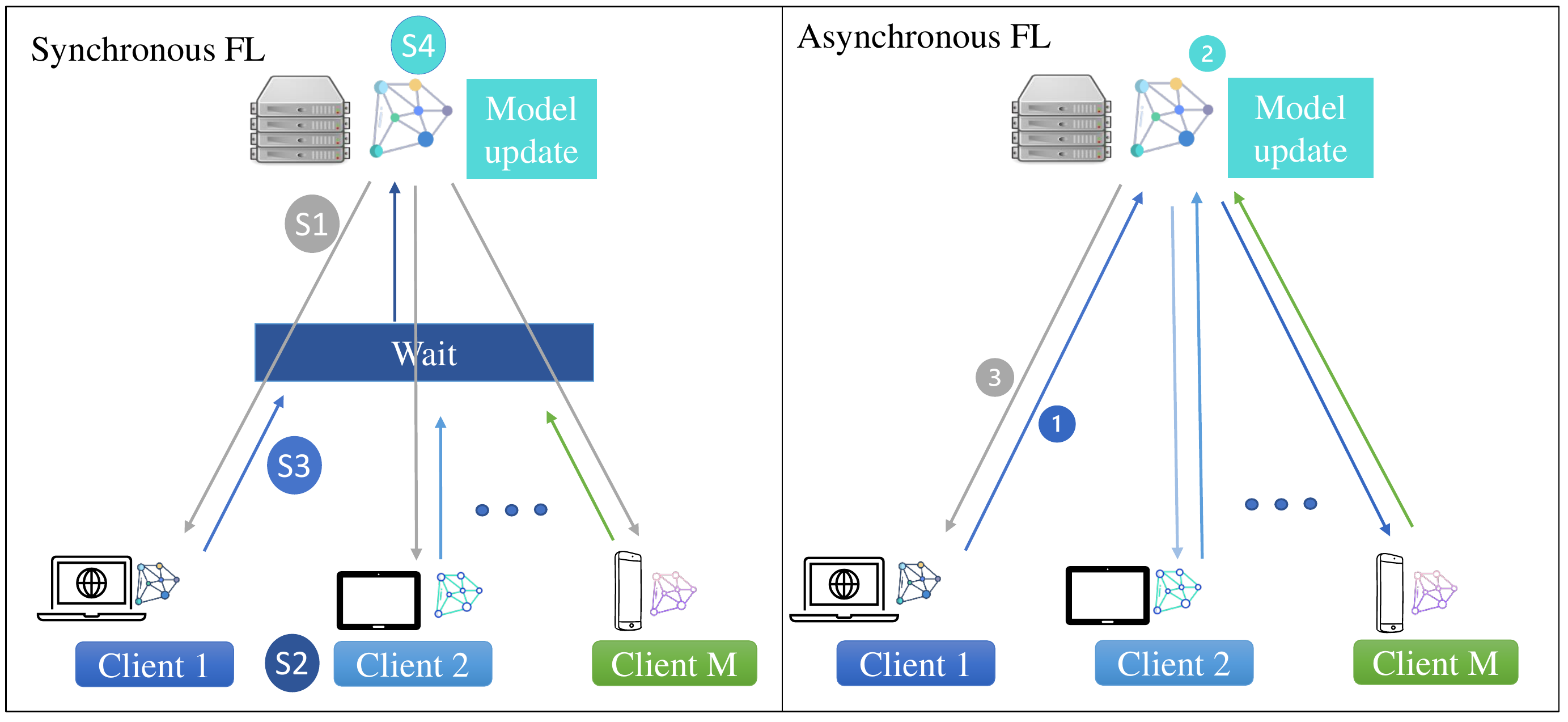}
	\caption{Synchronous vs Asynchronous FL.}
	\label{fig1}
\end{figure}

The global model is first initialized as $w_0$. Clients perform the local learning process as follows:
\begin{equation}
    w_t^m = w_t - \eta \nabla F_m(w_t), t=0,1,2,...,
\end{equation}
where $w_t^m$ is the local model of client $m$ at round $t$ after local learning, $w_t$ is the global model broadcast by the server at round $t$, $\eta$ represents the learning rate, and $\nabla F_m(w_t)$ signifies the gradient of the loss function $F_m(w_t)$. This takes place in step (S2) while the global broadcast model $w_t$ serves as the initial reference point for each client's local learning process.

Once the server receives the models from a predetermined number of clients or when a specified time limit elapses, the server proceeds to perform aggregation using the following process:
\begin{equation} \label{eq:global_update}
    w_{t+1} = \sum_{m=1}^M \alpha_m w_t^m,
\end{equation}
where $\alpha_m$ is the aggregation coefficient of client $m$, which is usually defined as $\alpha_m = \frac{|D_m|}{\sum_c |D_c|)}$. $|D_c|$ is the number of data samples at client $c$. Without loss of generality, we let all clients participate in each learning round as in equation (\ref{eq:global_update}). 

Based on the procedure mentioned above, the model of each client is synchronized with the global model following step (S1) in each round. Subsequently, in step (S2), the local models become different across clients after local learning. Step (S3) acts as a blocking operation, thus expanding the overall learning time. Lastly, in step (S4), the global model is updated by using equation (\ref{eq:global_update}).

\subsection{Asynchronous Federated Learning}
Under asynchronous settings, the server is not mandated to wait. Instead, the server initiates the aggregation process as soon as a model is received from a client. This approach also enables faster clients to proceed with local learning without waiting for slower clients. In SFL, a round is uniformly observed by both the server and all clients. However, faster clients perform more rounds in AFL than their slower counterparts. To keep track of the learning process, we will utilize the global aggregation number.
In order to differentiate the notation used in SFL, we employ the term ``iteration'' along with the symbols $i$ and $j$ to denote the global aggregation time in AFL. The aggregation process at the server is then performed as follows:
\begin{equation} \label{eq:asyn_globalupdate}
    w_{j+1} = \beta_{j} w_j  + (1-\beta_{j}) w_i^m,
\end{equation}
where $w_{j+1}$ is the global model in iteration $j+1$, $w_j$ represents the global model from the previous iteration $j$. The local model $w_i^m$ corresponds to the model obtained from client $m$ after local learning, utilizing the global model from iteration $i$ when client $m$ was reselected during iteration $j$. $\beta_{j} \in (0,1)$ denotes the aggregation coefficient. Following the aggregation process, the client $m$ that recently sent its model to the server receives the updated global model from the server. In Fig. \ref{fig1} (right), it is evident that only one client participates in the learning process during each iteration. Consequently, the subsequent local learning iteration continues based on the received updated global model, that is
\begin{equation} \label{eq:asyn_local}
    w_{j+1}^m = w_{j+1} - \eta \nabla F_m(w_{j+1}),
\end{equation}
where $w_{j+1}^m$ is the updated local model of client $m$. AFL follows a different approach where only one client receives the updated model during each global iteration. While client $m$ uploads its model to the server, other clients have the option to either continue their local computations or wait for the channel to become idle. This allows for concurrent local computations and efficient resource utilization in the AFL framework.

\subsection{Comparison}
\begin{figure}[!th]
	\centering
	\includegraphics[width=0.7\linewidth]{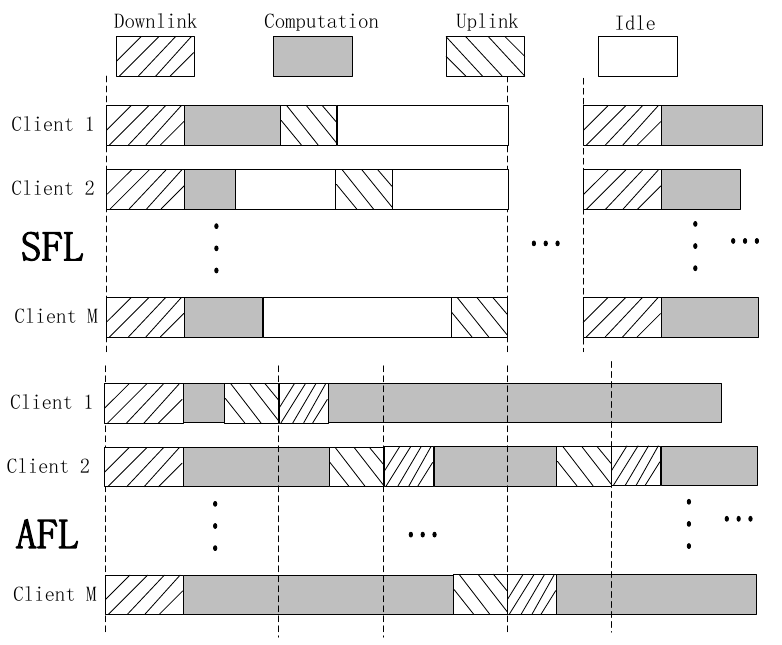}
	\caption{Time comparison between SFL and AFL}
	\label{fig2}
\end{figure}

Fig. \ref{fig2} shows the learning procedure of SFL (top) and AFL (bottom). There is extensive idle time in SFL while the learning continues in AFL. Based on the aforementioned details, we can infer that AFL is expected to learn faster than SFL as more computations are executed within the same time frame. To gauge the extent of AFL's speed, we analyze both modes in specific scenarios. First,  a homogeneous scenario is considered, where each client has identical computational capabilities. Let $\tau$ denote the computation time, which is the same for all clients. Another assumption is that a client can be scheduled to upload the local model again only when all other clients finished the model uploading. In terms of communication, the difference in channel conditions and power allocations are omitted here to facilitate the analysis. Thus, uploading time  $\tau^{u}$ in step (S3) is identical for all clients.

In step (S1), the time required to download the global model is assumed to be $\tau^{d}$. Hence, in SFL, the total completion time for one round, when employing time-division multiple access (TDMA), can be expressed as $\tau_{ho}^{syn} = \tau^d + \tau + M \cdot \tau^u$. Consequently, the global model receives updates after $\tau_{ho}^{syn}$ time has elapsed. On the other hand, in AFL, performing the same operation takes $\tau_{ho}^{asyn} = M \cdot \tau^u + M \cdot \tau^d + \tau$ time. Although AFL requires an additional $(M-1) \cdot \tau^d$ time compared to SFL to obtain the same global model, AFL updates the global model every $\tau^u + \tau^d$ time instead of waiting for $\tau^d + \tau + M \cdot \tau^u$ time as in SFL.

In the heterogeneous scenario, the computation capabilities of clients differ and the channel conditions vary. The computation time for the fastest client is assumed to be $\tau$, while the slowest client requires $a \cdot \tau$ time. Typically, computation is faster than communication, making communication the system's bottleneck. However, in situations where slow clients have parallel computation tasks, the value of $a$ can be substantial, causing $a \cdot \tau$ to exceed $M \cdot \tau^u$. Consequently, the completion time for one round is primarily dictated by the computation duration of the slowest client rather than communication. In SFL, the global model must wait for $\tau_{he}^{syn}=\tau^d + a \cdot \tau + M \cdot \tau^u$ time to get updated. This means that the faster clients must remain idle during this waiting period. In contrast, AFL achieves model updates within the range of $M\cdot \tau^d + \tau + M \cdot \tau^u \le \tau_{he}^{asyn} \le M\cdot \tau^d + a \cdot \tau + M \cdot \tau^u$ time, considering that all client models are uploaded when faster clients are scheduled first. Notably, in AFL, the global model is updated every $\tau^u + \tau^d$ time. Fig. \ref{fig2} illustrates this concept, where the vertical dashed line represents the aggregation time at the server. It is evident that the server performs aggregation more frequently in AFL.

The disparities between AFL and SFL contribute to AFL's faster learning pace, but they can also lead to model staleness among slow clients and potentially hinder the convergence of the global model. Consequently, it becomes crucial to identify the effective aggregation coefficient, $\beta$, and implement client scheduling during the learning process. Client scheduling plays a vital role in promoting convergence throughout the learning process while optimizing the aggregation coefficient aims to minimize the individual client's model staleness. These strategies work in tandem to ensure overall convergence and minimize the impact of model staleness in AFL.

\section{Proposed Algorithm}
In this section, we first utilize the SFL aggregation coefficient in AFL. Subsequently, we introduce an AFL algorithm designed to attain comparable learning performance to SFL. Lastly, we propose a client scheduling and model aggregation framework in AFL (CSMAAFL).

\subsection{SFL Aggregation Coefficient in AFL}
In SFL, client scheduling is unnecessary since all clients actively participate in model updates. Furthermore, the aggregation coefficient is determined by considering the relative number of samples present on each client, i.e.,
\begin{equation}
    \sum_{m=1}^m \alpha_m = \sum_{m=1}^M \frac{|D_m|}{\sum_c |D_c|)} = 1.
\end{equation}
Here, $\alpha_m$ represents the relative importance of the client $m$ in terms of its model's contribution. However, when using the relative number of samples on each client, denoted as $\alpha_m$, as the aggregation coefficient in AFL, the influence of initially selected clients diminishes as the iterations progress. Given a specific client scheduling sequence $\phi(1), \phi(2), \dots, \phi(M)$, where $\phi(i)$ indicates the client scheduled to upload its local model in iteration $i$, equation (\ref{eq:asyn_globalupdate}) can be expressed as follows:
\begin{equation}
\begin{aligned}
        w_{j+1} & = (1- \alpha_{\phi(j)})w_j + \alpha_{\phi(j)} w_i^{\phi(j)} \\
                & = (1- \alpha_{\phi(j)}) ((1- \alpha_{\phi(j-1)})w_{j-1} + \alpha_{\phi(j-1)} w_k^{\phi(j-1)})    \\
                & \quad + \alpha_{\phi(j)} w_i^{\phi(j)}, 
\end{aligned}
\end{equation}
where $\phi(j-1)$ represents the client scheduled in iteration $j-1$, and $k$ denotes the iteration when client $\phi(j-1)$ was scheduled to upload its model last time. Consequently, the aggregation coefficient for client $\phi(j-1)$ can be calculated as $\alpha_{\phi(j-1)}(1-\alpha_{\phi(j)})$. For the first client in the scheduling sequence, the aggregation coefficient is $\alpha_{\phi(1)}(1-\alpha_{\phi(2)})(1-\alpha_{\phi(3)})\cdots(1-\alpha_{\phi(j)})$. Since $\alpha$ falls within the range of $(0,1)$, the aggregation coefficient diminishes over time.

\subsection{Baseline}
To achieve comparable learning performance to SFL, AFL needs to adopt the same client scheduling strategy and aggregation coefficient. In AFL, a client is scheduled to upload its model again only when all other clients have finished uploading theirs. Additionally, faster clients are prioritized in the scheduling, allowing them to upload while slower clients are still performing computations. As for the aggregation weight $\beta$, it varies in each global iteration. In SFL, the relative model importance $\alpha_m$. To ensure that clients contribute equivalently as in SFL, the aggregation weight $\beta$ should also be calculated according to the contribution of each client $m$. This relationship is formulated in the following equation:
\begin{equation} \label{eq:syn_asyn}
    \sum_{m=1}^M \alpha_m w^m = w_{M+1} = \beta_{M}w_{M} + (1-\beta_{M}) w^{\phi(M)}.
\end{equation}
In Equation (\ref{eq:syn_asyn}), the left-hand side (LHS) corresponds to the global model after aggregation in SFL, while the right-hand side (RHS) represents the global model after iterating through all clients once in AFL. By examining Equation (\ref{eq:syn_asyn}), we know that $\beta_j$ is connected to both the iteration $j$ and the scheduled client $\phi(j)$.
\begin{equation} \label{eq:beta_calculation}
    \begin{aligned}
        w_M & = \beta_{M-1}w_{M-1}+(1-\beta_{M-1}) w^{\phi(M-1)} \\
        & = \beta_{M-1} (\beta_{M-2}w_{M-2}+(1-\beta_{M-2}) w^{\phi(M-2)}) \\
        & \quad + (1-\beta_{M-1}) w^{\phi(M-1)} \\
        & = \beta_1(\dots) + (1-\beta_1)w^{\phi(1)}.
    \end{aligned}
\end{equation}

From Equation (\ref{eq:beta_calculation}), we know that when the client scheduling $\phi(1), \phi(2), \dots, \phi(M)$ is predetermined, the only unknown parameters are $\beta_1, \beta_2, \dots, \beta_M$. On the other hand, $\alpha_1, \alpha_2, \dots, \alpha_M$ are known, and this knowledge allows us to formulate a set of $M$ non-linear equations. Non-linear equations typically have multiple solutions. By observing Equation (\ref{eq:syn_asyn}), we can ascertain that client $\phi(M)$ is selected during iteration $M$. An equation
\begin{equation}
\setlength{\abovedisplayskip}{3pt}
    \setlength{\belowdisplayskip}{3pt}
    \alpha_{\phi(M)} = 1-\beta_M,
\end{equation}
is formulated. Given that $\alpha_{\phi(M)}$ is a known value, we can solve for $\beta_M$. By considering Equation (\ref{eq:syn_asyn}) and Equation (\ref{eq:beta_calculation}), it becomes apparent that
\begin{equation}
    \alpha_{\phi(M-1)} = \beta_M(1-\beta_{M-1}).
\end{equation}
Consequently, we can proceed to solve for $\beta_{M-1}$. Following this approach, we can iteratively solve for $\beta_{M-2}, \beta_{M-3}$, and so on, until reaching $\beta_1$.

As outlined above, we have established a baseline for AFL to achieve comparable learning performance to SFL. This baseline entails the following requirements: a) clients are scheduled again for upload only when all other clients have been scheduled, b) client scheduling is predetermined prior to the learning process, and c) the global model is distributed to all clients every $M$ iterations. However, requirement a) results in the under-utilization of the computational capabilities of faster clients, which hampers the full potential of AFL. Furthermore, requirement b) imposes a relatively strong assumption, necessitating the server's knowledge of each client's computational capabilities beforehand. Lastly, requirement c) entails that clients halt local learning or discard their local learning models in favor of the global model.

\subsection{CSMAAFL: Client Scheduling and Model Aggregation in AFL  }
To fully leverage the advantages offered by AFL, we propose a client scheduling approach that incorporates a model aggregation scheme. This scheduling method takes into account both the computational capabilities of clients and the principle of fairness. When a client completes its local computation, it requests a time slot for uploading its updated local model. Subsequently, upon receiving the most recent aggregated global model, the client proceeds with learning for the next iteration. In situations where two clients complete their local computations simultaneously and apply for an uploading time slot, priority is given to the client with the older model. For instance, if clients $m$ and $n$ finish their local computations at the present time and aim to upload their updated local models during time slot $k$, and if the previous uploading slots for clients $m$ and $n$ are denoted as $m'$ and $n'$ respectively, then client $m$ will be prioritized if $(k - m') > (k - n')$.

The client scheduling approach described above is effective when there is relatively little variation in the computation capabilities among clients. However, there are two extreme scenarios that need to be considered. The first scenario arises when there are a few extremely fast clients, potentially operating at significantly accelerated speeds (e.g., $10$ times faster). The second scenario occurs when there are some excessively slow clients. To ensure that all clients have a fair opportunity to contribute to the global model, we employ a policy similar to the one outlined in \cite{fed_edge}. This policy allows clients with greater computation capabilities to perform more local iterations, thereby dedicating more time to the learning process. Conversely, clients with lower computation capabilities perform fewer local iterations, enabling them to spend less time on the task. By adopting this approach, we can maintain a balanced contribution from all clients, regardless of their varying computation speeds.

The client scheduling problem has been addressed, ensuring that each client has an equitable opportunity to upload their local updates. It needs to be addressed how to strike a balance between the current global model and the uploaded local models in each iteration. In equation (\ref{eq:asyn_globalupdate}), it can be observed that the contribution of client $m$ to the global model diminishes over time. Additionally, the difference between the current iteration and the iteration when client $m$ last uploaded its model, denoted as $j-i$, also plays a role. A smaller value of $j-i$ indicates a lower level of staleness. To account for this, we introduce the moving average $\mu_{ji}$, which captures the average value of $j-i$ over time. Let 
\begin{equation} \label{eq:asyn_beta}
    (1-\beta_j) w_i^m = \min(1, \frac{\mu_{ji}}{\gamma j(j-i)}) w_i^m,
\end{equation}
for equation (\ref{eq:asyn_globalupdate}), where $\gamma$ is a positive constant value. The term $\frac{1}{j}$ reflects the gradual decrease in the contribution of individual client models over time. The effect of staleness is represented by $\frac{\mu_{ji}}{j-i}$. When the learning starting point $i$ of a client $m$ is recent (i.e. when $j-i$ is small), the value of $\frac{\mu_{ji}}{j-i}$ is large, indicating a significant contribution from the individual client. As previously mentioned, extremely fast or slow clients are instructed to perform more or fewer local computations during their learning process, ensuring that every client has a comparable opportunity to access the channel for uploading updated local models. This approach results in only slight changes in $j-i$, leading to the value of $\frac{\mu_{ji}}{j-i}$ remaining close to 1. This helps maintain stability in the system while still accounting for the effects of staleness.
The complete algorithm is summarized in Algorithm \ref{al:asyn}.

\begin{algorithm}[]
\caption{Asynchronous Federated Learning with Client Scheduling and Model Aggregation} 
\begin{algorithmic}[1] \label{al:asyn}
\STATE  {\bf Initialization:} {\bf Server} initializes $w_0$ and broadcasts to all {\bf Clients}.
\WHILE{not converge}
\STATE {\bf Client:}\\
    \quad Receives the most recent aggregated global model. \\
    \quad Performs local computation as Eq.(\ref{eq:asyn_local}). \\
    \quad Applies for uploading time slot and uploads the aggregated model.
\STATE {\bf Server:} \\
    \quad Approves the first client $m$ requested the time slot.\\
    \quad Receives the local model from client $m$. \\
    \quad Performs aggregation by Eq. (\ref{eq:syn_asyn}) and Eq. (\ref{eq:asyn_beta}). \\
    \quad Sends the aggregated global model to client $m$.
\ENDWHILE
\end{algorithmic}
\end{algorithm}

\section{Simulation results}
In this section, we begin by outlining the simulation settings. Subsequently, we present the simulation results for MNIST and Fashion-MNIST datasets in both IID and non-IID scenarios. Finally, we analyze and discuss the impact of the constant $\gamma$ on the results.

A typical FL setting is considered for simulation here. The setup involves 100 clients connected to the server. In the case of SFL, all clients participate in the learning process during each round. However, for AFL, a client is waiting for its next upload only when all other clients have completed their current uploads. To simulate the heterogeneity in clients' computation capabilities, client selection is randomized in each trunk time, corresponding to the round time in SFL. Consequently, this random selection affects the values of $j-i$ and $\mu_{ji}$. Two public image datasets, i.e., MNIST and Fashion-MNIST, are used for simulation. MNIST consists of handwritten digit images, while Fashion-MNIST comprises images of Zalando's articles. Both datasets feature 10 classes, with 60,000 training images and 10,000 testing images. Under the IID scenario, the images are randomly allocated equally among the clients. However, in the non-IID scenario, each client is assigned two classes, resulting in approximately 600 training images per client. For the machine learning tasks, we employ Convolutional Neural Networks (CNN) with two convolutional layers, two max-pooling layers, and two fully connected layers. Given the complexity of the Fashion-MNIST images, the hidden layer sizes in the CNN for Fashion-MNIST are larger. The activation function for the last layer is the log softmax function, while ReLU is used in other layers. The learning rate $\eta$ is set as 0.01, and the local batch size is 5.  The constant $\gamma$ in equation (\ref{eq:asyn_beta}) can be considered a hyperparameter. A larger $\gamma$ value leads to smaller contributions from individual client models. To investigate the effect of $\gamma$, we set its value as 0.1, 0.2, 0.4, and 0.6, respectively.

Four simulation scenarios are considered, incorporating two datasets: MNIST and Fashion-MNIST, and two data distributions: IID and non-IID. In each scenario, we conduct simulations using both SFL and AFL approaches. The classical FedAvg algorithm is employed for SFL simulations, while our proposed CSMAAFL scheme is utilized for AFL simulations.

\begin{figure}[!th]
\setlength{\abovecaptionskip}{-0.2cm} 
	 \setlength{\belowcaptionskip}{-1cm}
	\centering
	\includegraphics[width=0.6\linewidth]{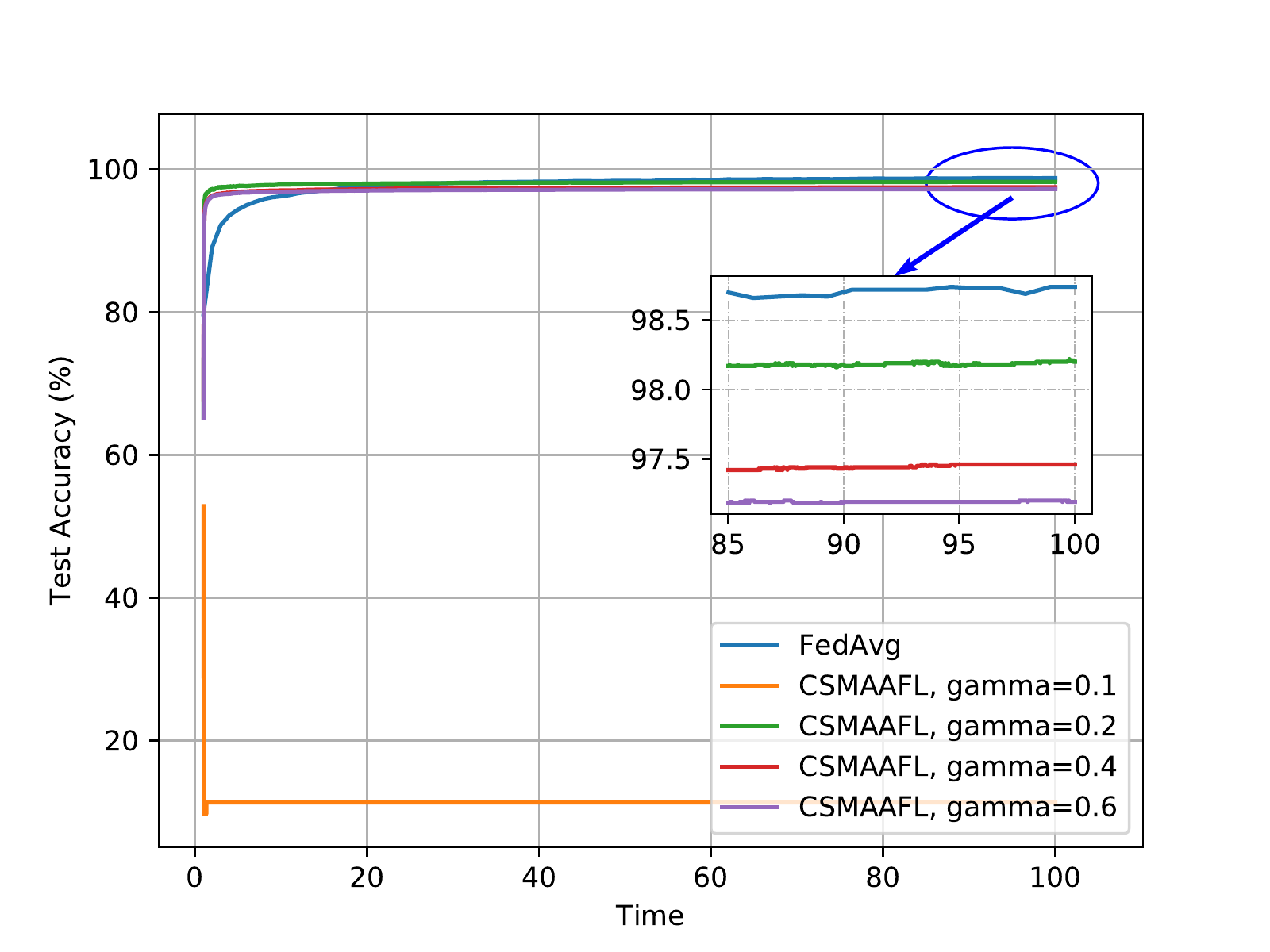}
	\caption{Scenario 1: MNIST IID}
	\label{fig:mnist_iid}
\end{figure}
In Fig. \ref{fig:mnist_iid}, the MNIST dataset with an IID data distribution is used. All schemes, except for CSMAAFL with $\gamma=0.1$, demonstrate comparable performance. This indicates that our proposed CSMAAFL approach converges and reaches a similar outcome as FedAvg when the value of $\gamma$ is properly tuned. 

\begin{figure}[!th]
	\centering
	\includegraphics[width=0.6\linewidth]{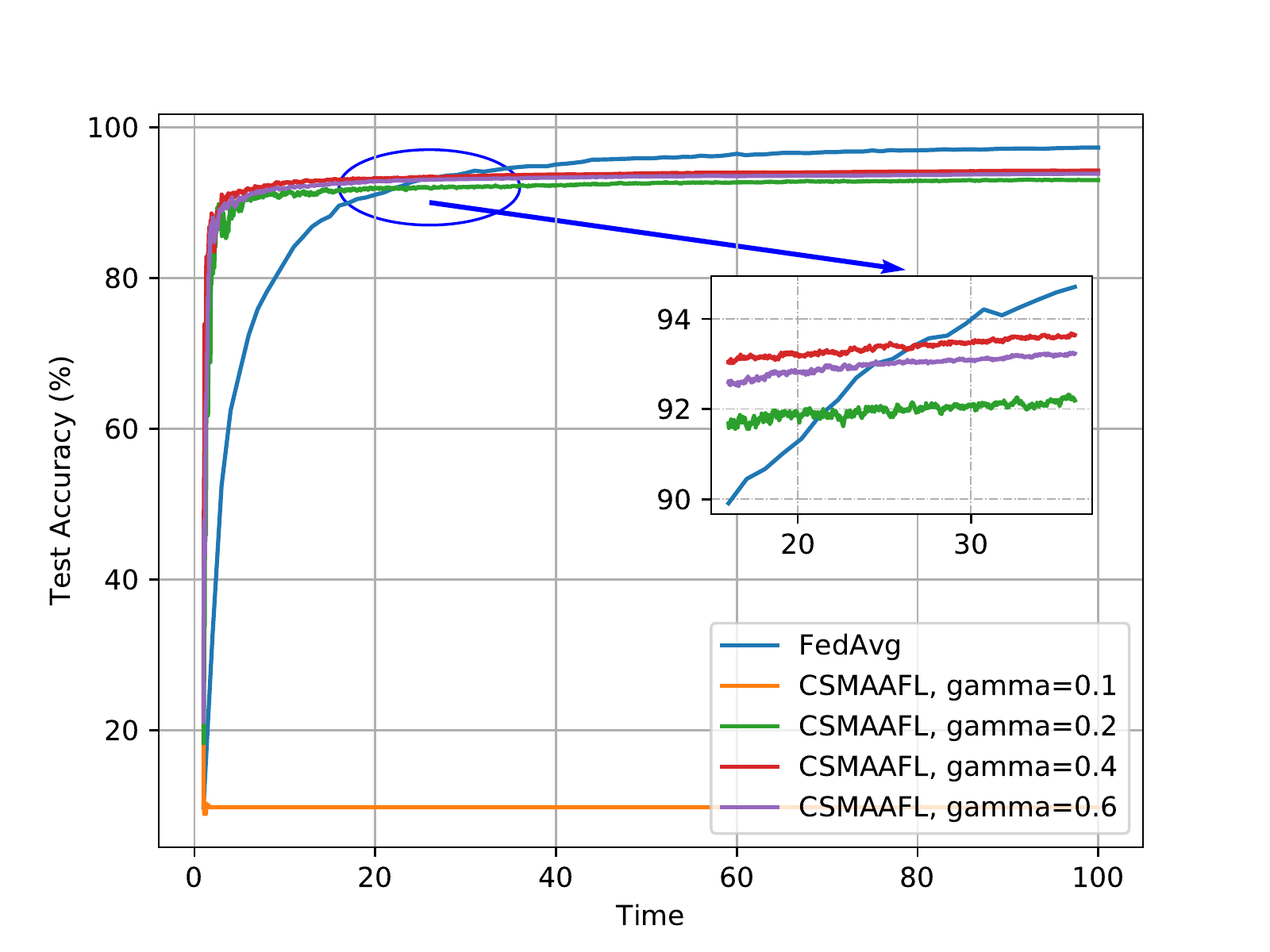}
	\caption{Scenario 2: MNIST non-IID}
	\label{fig:mnist_niid}
\end{figure}
In Fig. \ref{fig:mnist_niid}, the simulation is performed on the MNIST dataset with a non-IID data distribution. 
The FedAvg algorithm is starting to approach the performance of CSMAAFL after 25 relative time slots, which indicates the faster convergence advantage of the proposed algorithms.

\begin{figure}[ht]
  \centering
  \subfloat[Scenario 3: Fashion-MNIST IID]{\includegraphics[width=0.5\textwidth]{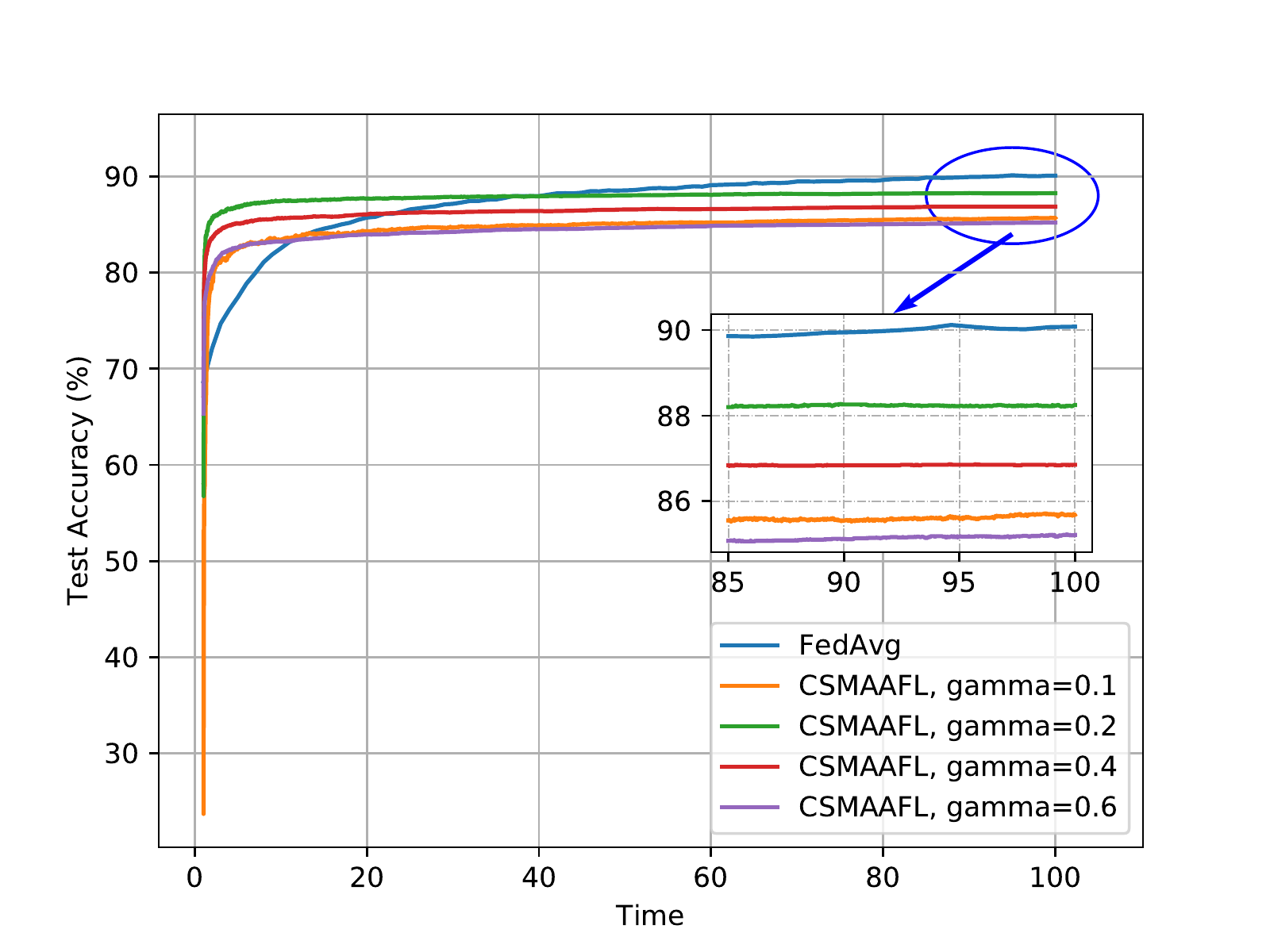}\label{fig:f1}}
  \hfill
  \subfloat[Scenario 4: Fashion-MNIST non-IID]{\includegraphics[width=0.5\textwidth]{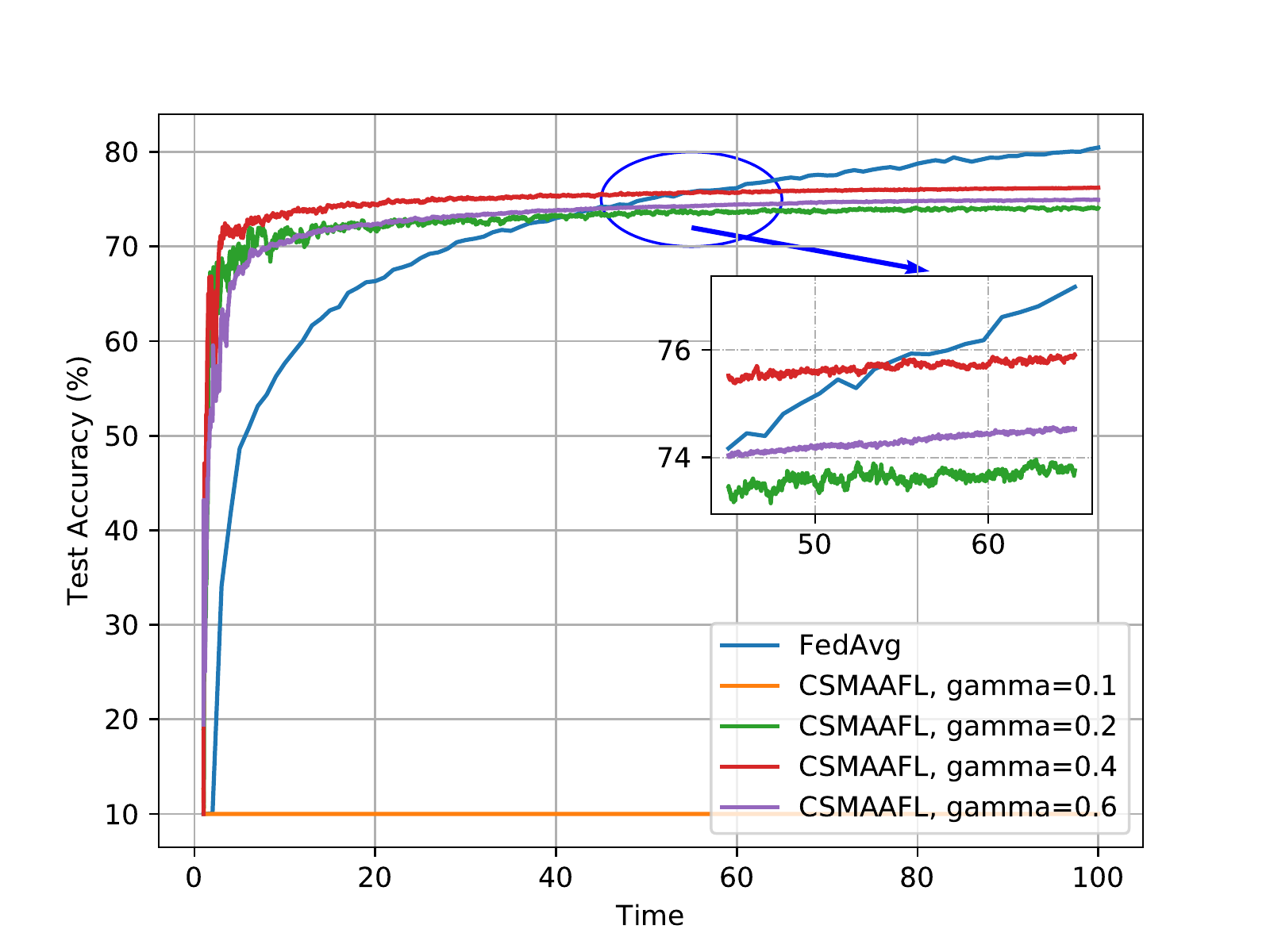}\label{fig:f2}}
  \caption{Fashion-MNIST}
  \label{fig:fmnist}
\end{figure}

The results for the Fashion-MNIST dataset with IID distribution are depicted in Fig. \ref{fig:fmnist}\subref{fig:f1}. Among the different $\gamma$ values, CSMAAFL with $\gamma=0.2$ exhibits the best performance, closely resembling the performance of the FedAvg algorithm.

In Fig. \ref{fig:fmnist}\subref{fig:f2}, it can be observed that CSMAAFL with $\gamma=0.6$ achieves a performance that closely matches that of FedAvg. However, it takes FedAvg 55 relative time slots to reach the same performance level as our proposed CSMAAFL scheme. This result once again demonstrates that our CSMAAFL scheme accelerates the learning performance during the initial stages while maintaining the overall learning performance.

In different scenarios, the choice of a constant value for $\gamma$ leads to varying effects. In the case of MNIST IID, MNIST non-IID, and Fashion-MNIST non-IID, a value of $\gamma=0.1$ results in random guessing. This occurs because the contribution of the individual client model is overly emphasized. On the other hand, for MNIST IID and Fashion-MNIST IID, the best performance is achieved with $\gamma=0.2$, while for MNIST non-IID and Fashion-MNIST non-IID, the optimal results are obtained with $\gamma=0.4$. By tuning $\gamma$, better learning performance can be achieved in CSMAAFL.

\section{Conclusions}
In this study, we introduced a client scheduling and model aggregation scheme for asynchronous federated learning. Our approach took into account both the computation capability and fairness of the clients in the scheduling process, while also addressing the issues of individual client contribution and model staleness in model aggregation. The results demonstrated that the proposed scheme can accelerate the federated learning process during the initial stages, while still achieving comparable performance to the synchronous algorithm.

\vspace{12pt}
\end{document}